\documentclass[12pt]{article}
\usepackage{epsfig,psfig, amssymb}
\def\be{\begin{equation}}
\def\ee{\end{equation}}
\def\bc{\begin{center}}
\def\ec{\end{center}}
\def\bea{\begin{eqnarray}}
\def\eea{\end{eqnarray}}
\def\dd{\displaystyle}
\def\nn{\nonumber}
\def\gappeq{\mathrel{\rlap {\raise.5ex\hbox{$>$}} {\lower.5ex\hbox{$\sim$}}}}
\def\lappeq{\mathrel{\rlap{\raise.5ex\hbox{$<$}} {\lower.5ex\hbox{$\sim$}}}}

\catcode`@=11
\def\marginnote#1{}
\newcount\hour
\newcount\minute
\newtoks\amorpm
\hour=\time\divide\hour by60
\minute=\time{\multiply\hour by60 \global\advance\minute by-\hour}
\edef\standardtime{{\ifnum\hour<12 \global\amorpm={am}%
        \else\global\amorpm={pm}\advance\hour by-12 \fi
        \ifnum\hour=0 \hour=12 \fi
        \number\hour:\ifnum\minute<10 0\fi\number\minute\the\amorpm}}
\edef\militarytime{\number\hour:\ifnum\minute<10 0\fi\number\minute}
\def\draftlabel#1{{\@bsphack\if@filesw {\let\thepage\relax
   \xdef\@gtempa{\write\@auxout{\string
      \newlabel{#1}{{\@currentlabel}{\thepage}}}}}\@gtempa
   \if@nobreak \ifvmode\nobreak\fi\fi\fi\@esphack}
        \gdef\@eqnlabel{#1}}
\def\@eqnlabel{}
\def\@vacuum{}
\def\draftmarginnote#1{\marginpar{\raggedright\scriptsize\tt#1}}
\def\draft{\oddsidemargin 0.0truein
        \def\@oddfoot{\sl preliminary draft \hfil
        \rm\thepage\hfil\sl\today\quad\militarytime}
        \let\@evenfoot\@oddfoot \overfullrule 3pt
        \let\label=\draftlabel
        \let\marginnote=\draftmarginnote
   \def\@eqnnum{(\theequation)\rlap{\kern\marginparsep\tt\@eqnlabel}%
\global\let\@eqnlabel\@vacuum}  }
\catcode`@=12
\begin{document}
\begin{titlepage}
\vspace*{-1cm}
\phantom{hep-ph/0402155} 
\hfill{DFPD-04/TH/05}

\hfill{CERN-PH-TH/2004-027}

\hfill{ROMA1-TH/1368-04}
\vskip 1.0cm
\begin{center} {\Large \bf Can Neutrino Mixings Arise from the \\ 
\vskip 0.2cm 
Charged Lepton Sector? }
\end{center}
\vskip 1.3  cm

\begin{center} {\bf \large Guido Altarelli}~\footnote{e-mail address: guido.altarelli@cern.ch},
\\
\vskip .1cm CERN, Department of Physics,  Theory Division, 
\\  CH-1211 Gen\`eve 23, Switzerland
\\
\vskip .3cm {\bf \large Ferruccio Feruglio}~\footnote{e-mail address: feruglio@pd.infn.it},
\\
\vskip .1cm Dipartimento di Fisica `G.~Galilei', Universit\`a di Padova and
\\  INFN, Sezione di Padova, Via Marzolo~8, I-35131 Padova, Italy
\\
\vskip .3cm {\bf \large Isabella Masina}~\footnote{e-mail address: isabella.masina@roma1.infn.it}
\\
\vskip .1cm Enrico Fermi Center, Via Panisperna 89/A, I-00184 Roma, Italy and 
\\  INFN, Sezione di Roma, P.le A. Moro 2, I-00185 Roma, Italy

\end{center}
\vskip 1.5cm
\begin{abstract} 
\noindent The neutrino mixing matrix $U$ is in general of the form $U=U_e^\dagger U_{\nu}$, where $U_e$ arises from the
diagonalization of charged leptons and $U_{\nu}$ is from the neutrino sector. We discuss the possibility that $U_{\nu}$ is
nearly diagonal (in the lagrangian basis) and the observed mixing arises with good accuracy from $U_e$. We find that the fact that, 
in addition
to the nearly maximal atmospheric mixing angle $\theta_{23}$, the solar angle $\theta_{12}$ is definitely also large while at
the same time the third mixing angle $\theta_{13}$ is small, makes the construction of a natural model of this sort considerably 
more complicated. We present an example of a natural model of this class. 
We also find that the case that $U_\nu$ is exactly of the bimixing type is severely constrained 
by the bound on $\theta_{13}$ but not excluded. 
We show that planned experimental searches for $\theta_{13}$ could have a strong impact on bimixing models.

\end{abstract}
\end{titlepage}
\setcounter{footnote}{0}
\vskip2truecm

%
%

\section{Neutrino Mixings from the Charged Lepton \\Sector}

The observed neutrino mixing matrix $U=U_e^\dagger U_{\nu}$, in the limit of vanishing $\sin{\theta_{13}}=s_{13}$, has the
approximate form:
\be U=
\left[\matrix{
c&s&0\cr s/\sqrt{2}&-c/\sqrt{2}&1/\sqrt{2}\cr -s/\sqrt{2}&c/\sqrt{2}&1/\sqrt{2}} 
\right]~~~~~, 
\label{Uobs}
\ee
where $s$ and $c$ stand for $\sin{\theta_{12}}$ and $\cos{\theta_{12}}$ respectively and we took the atmospheric angle
$\theta_{23}$ as exactly maximal. The effective mass matrix of light neutrinos is in general given by:
\be
m_\nu= U^* m_{\nu}^{diag}U^\dagger~.~~~~~~\label{mnu}\\
\ee
Starting from the lagrangian basis, where all symmetries of the theory are specified, we want to investigate whether it is possible to obtain the observed mixings in a natural way from the diagonalization of the charged
lepton mass matrix by $U_e$ while $U_{\nu}$ is nearly diagonal. The possible deviations from maximal $\theta_{23}$ and from $s_{13}=0$ can be
omitted in eq. (\ref{Uobs}) and attributed to small effects from $U_{\nu}$ that will be in general not exactly zero. One
might think that given the rather symmetric role of $U_e$ and $U_{\nu}$ in the formula $U=U_e^\dagger U_{\nu}$ one way or the
other should be equivalent. But we will show that this is not so. Actually now that we know that also the solar angle
$\theta_{12}$ is large, this tends to clash with a small $\theta_{13}$, in the case of mixings dominated by $U_e$. 

In terms of $U_e$ the charged lepton mass matrix $m_e$ (defined as $\bar R m_eL$ from right-handed ($R$) and left-handed
($L$) charged lepton fields in the lagrangian basis) can be written as:
\be
m_e=V_e m_e^{diag} U_e^\dagger~~.~~~~~~\label{me}\\
\ee
Indeed $L_{diag}= U_e L$ and $R_{diag}=V_e R$ are the transformations between the lagrangian and the mass basis for the $R$ and $L$
fields. Assuming that $U\sim U_e^\dagger$, given that $m_e^{diag}=Diag[m_e,m_\mu,m_\tau]$ we can write:
\be
m_e=V_e m_e^{diag} U= V_e \left[\matrix{
cm_e&sm_e&0\cr s/\sqrt{2}m_\mu&-c/\sqrt{2}m_\mu&m_\mu/\sqrt{2}\cr -s/\sqrt{2}m_\tau&c/\sqrt{2}m_\tau&m_\tau/\sqrt{2}} 
\right]~~~~~.\label{meV}
\ee
We will come back later on the matrix $V_e$ that determines the right-handed mixings of charged leptons. For the time being
it is already interesting to consider the matrix $m_e^\dagger m_e$ which is completely fixed by $U_e$: 
\be
m_e^\dagger m_e=U_e(m_e^{diag})^2 U_e^\dagger~~.
\label{me2}\\
\ee
Neglecting for simplicity the electron mass, we find, for $U_e^\dagger=U$:
\be
m_e^\dagger m_e= U^\dagger(m_e^{diag})^2 U = \frac{1}{2}(m_\tau^2+m_\mu^2)\left[\matrix{
s^2&-cs&-s(1-2\lambda^4)\cr -cs&c^2&c(1-2\lambda^4)\cr
-s(1-2\lambda^4)&c(1-2\lambda^4)&1} 
\right]~~~~~,\label{me22}
\ee
where we defined
\be
\frac{m_\tau^2-m_\mu^2}{m_\tau^2+m_\mu^2} = 1-2\lambda^4~~~~~\label{eta}\\
\ee
so that approximately $\lambda^4 \sim m_{\mu}^2/m_{\tau}^2$. The problem with this expression for $m_e^\dagger m_e$ is that all
matrix elements are of the same order and the vanishing of $s_{13}$ as well as the hierarchy of the eigenvalues arise from
precise relations among the different matrix elements. For example, the result $s_{13}=0$ is obtained because the
eigenvector with zero eigenvalue is of the form $e_1=(c,s,0)^T$ and the crucial zero is present because 
the first two columns are proportional in eq. (\ref{me22}). These features are more difficult to implement in a
natural way than matrices with texture zeros or with a hierarchy of matrix elements. Only if the solar angle $\theta_{12}$
is small, that is $s$ is small, then the first row and column are nearly vanishing and $s_{13}$ is automatically small.

Consider, for comparison, the case where we do not make the
hypothesis that all the mixings are generated by the charged leptons, but rather that $U_e \sim 1$. 
To make the comparison more direct, let us assume
that the neutrino mass spectrum is of the normal hierarchy type with dominance of
$m_3$: $m_\nu^{diag}\sim m_3 Diag[0,\xi^2,1]$, where $\xi^2=m_2/m_3$ is small and $m_1$ is neglected. In this case, the
effective light neutrino mass matrix is given by (note the crucial transposition of $U$, 
which in eq. (\ref{Uobs}) is real,  with respect to eq. (\ref{me22})):
\be
m_\nu = U^*  m_\nu^{diag} U^\dagger \sim \frac{m_3}{2}\left[\matrix{
s^2\xi^2&-cs\xi^2/\sqrt{2}&cs\xi^2/\sqrt{2}\cr -cs\xi^2/\sqrt{2}&(1+c^2\xi^2)/2&(1-c^2\xi^2)/2\cr
cs\xi^2/\sqrt{2}&(1-c^2\xi^2)/2&(1+c^2\xi^2)/2} 
\right]~~~~~.\label{menu}
\ee
In this case, no matter what the value of $s$ is, the first row and column are of order $\xi^2$. By replacing terms of
order $\xi^2$ by generic small terms of the same order, $s_{13}$ remains of order $\xi^2$. We can also replace the
terms of order 1 in the 23 sector by generic order 1 quantities provided that we have a natural way of guaranteeing that the
subdeterminant 23 is suppressed and remains of order $\xi^2$. As well known this suppression can be naturally induced
through the see-saw mechanism either by dominance of a single right-handed Majorana neutrino \cite{dominance} or by a lopsided \cite{lopsided} neutrino Dirac
matrix. Natural realizations of this strategy have been constructed, for example, in the context of U(1)$_F$ flavour models \cite{lopsu1}.

We now come back to the expression for the charged lepton mass matrix $m_e$ in eq. (\ref{meV}) where the matrix $V_e$ appears.
This matrix describing the right-handed mixings of charged leptons is not related to neutrino mixings. In minimal SU(5) the
relation $m_e=m_d^T$ holds between the charged lepton and the down quark mass matrices. In this case $V_e$ describes the
left-handed down quark mixings: $V_e=U_d$. The CKM matrix, as well known, is given by $V_{CKM}=U_u^\dagger U_d$. Given that
the quark mixing angles are small, either both $U_u$ and $U_d$ are nearly diagonal or they are nearly equal. Thus one
possibility is that $U_d$ is nearly diagonal. In this case, for $V_e=U_d$, $m_e$ is approximately given by eq. (\ref{meV}) with $V_e \sim 1$.
Neglecting the electron mass and setting $\lambda^2=m_{\mu}/m_{\tau}$ we obtain:
\be
m_e\approx m_e^{diag} U= m_{\tau}\left[\matrix{
0&0&0\cr s/\sqrt{2}\lambda^2&-c/\sqrt{2}\lambda^2&\lambda^2/\sqrt{2}\cr -s/\sqrt{2}&c/\sqrt{2}&1/\sqrt{2}} 
\right]~~~~~.\label{meV1}
\ee
This matrix is a generalization of lopsided models with all three matrix elements in the third row of order 1 (unless $s$ is
small: for small solar angle we go back to the situation of normal lopsided models). We recall that lopsided models
with the 23 and 33 matrix elements of order 1 provide a natural way to understand a large 23 mixing angle. In fact from the
matrix relation
\be
\left[\matrix{
0&0&0\cr 0&0&0\cr 0&s_{23}&c_{23}\cr} 
\right]\left[\matrix{
1&0&0\cr 0&c_{23}&s_{23}\cr 0&-s_{23}&c_{23}\cr} 
\right]= \left[\matrix{
0&0&0\cr 0&0&0\cr 0&0&1\cr} 
\right]~~~~~,\label{lop}
\ee
we see that in lopsided models one automatically gets a large 23 mixing from $U_e$. In the generalized case of
eq. (\ref{meV1}), while the natural prediction of a large 23 mixings remains, the relation $s_{13}=0$ does not arise
automatically if the entries of the matrix are replaced by generic order 1 terms in the third row and of order $\lambda^2$
in the second row. If we call $v_3$ the 3-vector with components of order 1 in the third row and $\lambda^2 v_\lambda$ the
vector of the second row, we can easily check that to obtain $s_{13}=0$ it is needed that $v_\lambda v_3 =0$ and also that
$v_\lambda$ is orthogonal to a vector of the form $(c,s,0)$.

In democratic models all matrices $U_u$, $U_d$, $U_e$ are nearly equal and the smallness of quark mixings
arises from a compensation between $U_u^\dagger$ and $U_d$. This sort of models correspond, for $V_e=U_d$, to $V_e=U_e=U^\dagger$ and a symmetric
matrix $m_e$: $m_e=U^\dagger m_e^{diag} U$. In this case we obtain a matrix exactly equal to that in eq. (\ref{me22}) for $m_e^\dagger
m_e$ except that squared masses are replaced by masses. As discussed in the case of eq. (\ref{me22}), 
we need fine-tuning in order to reproduce the observed hierarchy of mass and to obtain
$s_{13}=0$ unless the solar angle $s$ is small. Note in fact, that in the democratic model of \cite{fldem},
the vanishing of $s_{13}$ is only accommodated but not predicted.

\section{A Natural Class of Models}

We now attempt to identify a set of conditions that make possible the
construction of an explicit model where the mixing in the lepton sector
is dominated by the charged lepton contribution.
One obvious condition is a dynamical or a symmetry principle that forces
the light neutrino mass matrix to be diagonal in the lagrangian basis.
The simplest flavour symmetries cannot fulfill this requirement in a 
simple way.
For instance, a U(1) symmetry can lead to a nearly diagonal neutrino mass
matrix, of the form:
\be 
m_\nu=
\left[\matrix{
\xi^{2p}&\xi^{p+1}&\xi^p\cr 
\xi^{p+1}&\xi^2&\xi\cr 
\xi^p&\xi&1} 
\right] m~~~~~, 
\label{mnund}
\ee
where $\xi<1$ is a U(1) breaking parameter, $p\ge 1$
and all matrix elements are defined up to unknown order one coefficients.
The problem with this matrix is that the ratio between the solar and the 
atmospheric squared mass differences, close to 1/40, is approximately
given by $\xi^4$ and, consequently,
a large atmospheric mixing angle is already induced by $m_\nu$ itself.
If we consider a discrete symmetry like $S_3$, $m_\nu$ can
be of the general form:
\be 
m_\nu=
\left[\matrix{
1&0&0\cr 
0&1&0\cr 
0&0&1} 
\right] m +
\alpha \left[\matrix{
0&1&1\cr 
1&0&1\cr 
1&1&0} 
\right] m
~~~, 
\label{mnus3}
\ee
where $\alpha$ is an arbitrary parameter. In this case we need 
both the extra assumption $|\alpha|\ll 1$ and a specific symmetry breaking 
sector to lift the mass degeneracy \cite{fldem}.
A stronger symmetry like O(3) removes from eq. (\ref{mnus3}) the non-diagonal
invariant, but requires a non-trivial symmetry breaking sector
with a vacuum alignment problem in order to keep the
neutrino sector diagonal while allowing large off-diagonal terms
for charged leptons \cite{barbierio3}. 

A simple, though not economical, possibility to achieve a diagonal
neutrino mass matrix, is to introduce three independent U(1) symmetries,
one for each flavour: 
F=U(1)$_{F_1}\times$U(1)$_{F_2}\times$U(1)$_{F_3}\times$..., where F denotes the flavour symmetry group.
The Higgs doublet giving mass to the up-type quarks is neutral under F.
Each lepton doublet is charged under a different U(1) factor, with the same
charge +1. In the symmetric phase all neutrinos are exactly massless.
Flavour symmetry breaking is obtained by non-vanishing vacuum expectation 
values (VEVs) of three 
flavon fields, also charged under a separate U(1) factor, with charge -2. 
In this way only diagonal neutrino mass terms are induced. 
If the VEVs in the flavon sector
are similar, we expect neutrino masses of the same order, and 
the observed hierarchy between the solar and atmospheric squared mass
differences requires a modest adjustment of the flavon vev's and/or of the coefficients of 
the lepton violating operators.

A second condition can be identified by considering a mass matrix 
for the charged leptons which is very close, but slightly more 
general than the one of eq. (\ref{meV1}):
\be
m_e=\left[\matrix{
O(\lambda^4)&O(\lambda^4)&O(\lambda^4)\cr 
x_{21}\lambda^2& x_{22}\lambda^2&O(\lambda^2)\cr 
x_{31}&x_{32} &O(1)} 
\right] m~~~~~,
\label{meV2}
\ee
where $x_{ij}$ $(i=2,3)$ $(j=1,2)$ is a matrix of order one coefficients
with vanishing determinant:
\be
x_{21} x_{32} - x_{22} x_{31}=0~~~~~.
\label{det0}
\ee
The eigenvalues of $m_e$ in units of $m$ are of order 1, 
$\lambda^2$ and $\lambda^4$, as required by the charged lepton masses.
Moreover, the eigenvalue of order $\lambda^8$ of $m_e^\dagger m_e$ 
has an eigenvector:
\be
(c,s,O(\lambda^4))~~~~~~\frac{s}{c}=-\frac{x_{31}}{x_{32}}+O(\lambda^4)~~~~~.
\ee
In terms of the lepton mixing matrix $U=U_e^\dagger$, this means 
$\theta_{13}=O(\lambda^4)$ and $\theta_{12}$ large, if 
$x_{31}\approx x_{32}$. When the remaining, unspecified parameters in $m_e$
are all of order one, also the atmospheric mixing angle $\theta_{23}$
is large. 
Notice that, by neglecting $O(\lambda^4)$ terms, the following relation
holds for the mass matrix $m_e/m$ (cfr. eq. (\ref{lop})):
\be
\left[\matrix{
0&0&0\cr 
x_{21}\lambda^2& x_{22}\lambda^2&O(\lambda^2)\cr 
x_{31}&x_{32} &O(1)} 
\right]
\left[\matrix{
c&-s&0\cr 
s&c&0\cr 
0&0&1} 
\right]=
\left[\matrix{
0&0&0\cr 
0& \sqrt{x_{21}^2+x_{22}^2}\lambda^2&O(\lambda^2)\cr
0&\sqrt{x_{31}^2+x_{32}^2} &O(1)} 
\right]~~~~~,
\label{meV3}
\ee
where
\be
\frac{s}{c}=-\frac{x_{31}}{x_{32}}~~~~~.
\ee
Therefore the natural parametrization of the unitary matrix
$U_e$ that diagonalizes $m_e^\dagger m_e$ in this approximation is:
\be
U_e={U^e}_{12} {U^e}_{23}~~~~~,
\ee
where $U^e_{ij}$ refers to unitary transformation in the $ij$ plane.
Using $U=U_e^\dagger$, we automatically find the leptonic mixing matrix 
in the standard parametrization $U=U_{23} U_{13} U_{12}$ (neglecting phases),
with $U_{23}={U^e}_{23}^\dagger$, $U_{13}=1$ and $U_{12}={U^e}_{12}^\dagger$.
Had we used the standard parametrization also for $U_e$, we would have
found three non-vanishing rotation angles $\theta^e_{ij}$ with non-trivial
relations in order to reproduce $\theta_{13}=0$.

This successful pattern of $m_e$, eq. (\ref{meV2}), has two features. The first one
is the hierarchy between the rows. It is not difficult to obtain this
in a natural way. For instance, we can require a U(1) flavour symmetry
acting non-trivially only on the right-handed charged leptons, thus
producing the required suppressions of the first and second rows.
The second one is the vanishing determinant condition of eq. (\ref{det0}).
We can easily reproduce this condition by exploiting a see-saw
mechanism operating in the charged lepton sector.

To show this we add to the field content
of the standard model additional vector-like fermion pairs 
$(L_a,L^c_a)$ $(a=1,...n)$ of SU(2) doublets, with hypercharges 
$Y=(-1/2,+1/2)$. The Lagrangian in the charged lepton sector
reads:
\be
{\cal L}={kinetic~terms}
+ \eta_{ij} e^c_i l_j h_d 
+ \lambda_{ia} e^c_i L_a h_d 
+ \mu_{aj} L^c_a l_j 
+ M_a L^c_a L_a + h.c.
\label{lag}
\ee
where $l_i$ and $e^c_i$ $(i=1,2,3)$ are the standard model leptons,
doublet and singlet under SU(2), respectively, and $h_d$ denotes
the Higgs doublet. We assume a diagonal mass matrix for the extra
fields. We expect $M_a,\mu_{aj}\gg \langle h_d\rangle$
and in this regime there are heavy fermions that can be
integrated out to produce a low-energy effective Lagrangian.
The heavy combinations are $L^c_a$ and 
\be
L_a+\frac{\mu_{aj}}{M_a} l_j~~~~~(a=1,...n)~~~.
\label{heavy}
\ee
These fields are set to zero by the equations of motion
in the static limit and we should express all remaining fermions
in term of the three combinations that are orthogonal to those
in eq. (\ref{heavy}), and which we identify with the light degrees of
freedom. To illustrate our point it is sufficient to work in the regime
$\vert\mu_{aj}\vert<\vert M_a\vert$ and expand the relevant quantities
at first order in $\vert\mu_{aj}/M_a\vert$.
To this approximation
the light lepton doublets $l'_i$ are:
\be
l'_i=l_i-\frac{\mu_{ai}}{M_a} L_a~~~~~.
\ee
The effective lagrangian reads:
\be
{\cal L}={kinetic~terms}
+ (\eta_{ij}-\frac{\lambda_{ia}\mu_{aj}}{M_a})~ e^c_i l'_j h_d 
+ h.c.~~~~~,
\label{lagr}
\ee
and the mass matrix for the charged leptons is:
\be
m_e=\langle h_d\rangle(\eta_{ij}-
\frac{\lambda_{ia}\mu_{aj}}{M_a})~~~~~~.
\ee
This result is analogous to what obtained in the neutrino sector
from the see-saw mechanism. There is a term in $m_e$
coming from the exchange of the heavy fields $(L_a,L^c_a)$, which
play the role of the right-handed neutrinos, and there is another
term that comes from a single operator and that cannot be interpreted
as due to the exchange of heavy modes. In the regime 
$1>\vert\mu/M\vert>\vert\eta/\lambda\vert$ the ``see-saw'' contribution dominates.
Moreover, if the lower left block in $m_e$
is dominated by a single exchange, for instance by $(L_1,L^c_1)$,
then
\be
\left[\matrix{
{m_e}_{21}&{m_e}_{22}\cr 
{m_e}_{31}&{m_e}_{32}}
\right]=\frac{\langle h_d\rangle}{M_1}
\left[\matrix{
\lambda_{21}\mu_{11}&\lambda_{21}\mu_{12} \cr 
\lambda_{31}\mu_{11}&\lambda_{31}\mu_{12}}
\right]~~~~~,
\ee
and the condition of vanishing determinant in eq. (\ref{det0})
is automatically satisfied.

Additional vector-like leptons are required by several extensions
of the standard model. For instance, a grand unified theory based
on the E$_6$ gauge symmetry group with three generations
of matter fields described by three 27 representations of E$_6$,
includes, beyond the standard model fermions, two SU(5) singlets
and an SU(5) vector-like $(5,\bar{5})$ pair per each generation.
In such a model a ``see-saw'' mechanism induced by the exchange
of heavy $(5,\bar{5})$ fields is not an option, but a necessary
ingredient to recover the correct number of light degrees of freedom.
We should still show that it is possible to combine the above 
conditions in a natural and consistent framework. In the Appendix A
we present, as an existence proof, a supersymmetric 
SU(5) grand unified model possessing a flavour symmetry 
F=U(1)$_{F_0}\times$U(1)$_{F_1}\times$U(1)$_{F_2}\times$U(1)$_{F_3}$.
The first U(1)$_{F_0}$ factor is responsible for the hierarchy of masses
and mixing angles in the up-type quark sector as well as for the
hierarchy between the rows in the charged lepton mass matrix. 
The remaining part of F guarantees a diagonal
neutrino mass matrix and, at the same time, dominance of a single
heavy $(5,\bar{5})$ pair in the lower left block of $m_e$.
Notice that, at variance with most of the other existing models
\cite{afrev}, this framework
predicts a small value for $\theta_{13}$, of order $\lambda^4$
which is at the border of sensitivity of future neutrino 
factories. 


\section{Corrections to Bimixing from $U_e$}

Even when the neutrino mass matrix $U_\nu$ is not diagonal in the lagrangian 
basis, the contribution
from the charged lepton sector can be relevant or even crucial to
reproduce the observed mixing pattern. An important example arises if 
the neutrino matrix $U_\nu$ instead of being taken as nearly diagonal, 
is instead assumed of a particularly simple form, like for bimixing: 
\be 
U_\nu=
\left[\matrix{
1/\sqrt{2}&1/\sqrt{2}&0\cr 1/2&-1/2&1/\sqrt{2}\cr -1/2&1/2&1/\sqrt{2}} 
\right]~~~~~. 
\label{Ubim}
\ee
This configuration can be obtained, for instance,
in inverse hierarchy models with a $L_e-L_\mu-L_\tau$ U(1) symmetry, 
which predicts maximal $\theta_{12}^\nu$, large $\theta_{23}^\nu$, vanishing $\theta_{13}^\nu$
and $\Delta m^2_{sol}=0$. 
After the breaking of this symmetry, the degeneracy between the first two neutrino 
generations is lifted and the small observed value of $\Delta m^2_{sol}$ can be easily reproduced.
Due to the small symmetry breaking parameters,
the mixing angles in eq. (\ref{Ubim}) also receive corrections, 
whose magnitude turns out \cite{lms2} to be controlled by $\Delta m^2_{sol}/\Delta m^2_{atm}$: 
$\theta_{13}^\nu \lesssim 1 - \tan^2 \theta_{12}^\nu \sim \Delta m^2_{sol}/(2\Delta m^2_{atm}) \sim 0.01$. 
These corrections are too small to account for the measured value of the solar angle. 
Thus, an important contribution from $U_e$ is necessary to reconcile bimixing with observation.

In this section we will reconsider the question of whether the observed 
pattern can result from the corrections induced by the charged lepton sector. 
Though not automatic, this appears to be at present a rather natural possibility \cite{chlept, King} 
- see also the recent detailed analysis of Refs. \cite{Andrea} and \cite{Frampton}.
Our aim is to investigate the impact of planned experimental improvements, 
in particular those on $|U_{e3}|$, on bimixing models.
To this purpose it is useful to adopt a convenient parametrization 
of mixing angles and phases.
Let us define
\be
\tilde U~=~ 
\left(\matrix{1&0&0 \cr 0&c_{23}&s_{23}\cr0&-s_{23}&c_{23}     } 
\right)
\left(\matrix{c_{13}&0&s_{13}e^{i\delta} \cr 0&1&0\cr -s_{13}e^{-i\delta}&0&c_{13}     } 
\right)
\left(\matrix{c_{12}&s_{12}&0 \cr -s_{12}&c_{12}&0\cr 0&0&1     } 
\right)~~~,
\label{ufi}
\ee 
where all the mixing angles belong to the first quadrant and $\delta$ to $[0,2 \pi]$.
The standard parameterization for $U$ reads: $U = \tilde U \times$ a diagonal $U(3)$ matrix 
accounting for the two Majorana phases of neutrinos (the overall phase is not physical).
Since in the following discussion we are not interested in the Majorana phases, 
we will focus our attention on $\tilde U$. 

It would be appealing to take the parameterization (\ref{ufi}) separately for $U_e$ and $U_\nu$, 
by writing $s_{12}$, $s^e_{12}$, $s^\nu_{12}$ etc to distinguish the mixings of the 
$U$, $U_e$ and $U_\nu$ matrices, respectively. 
However, as discussed in the Appendix B, even disregarding Majorana phases,
$U$ is not just determined in terms of $\tilde U_e$ and $\tilde U_\nu$, with the latter 
defined to be of the form (\ref{ufi}). 
The reason is that, by means of field redefinitions $U_e$ and $U_\nu$ can be separately but 
{\it not simultaneously} written respectively 
as $\tilde U_e$ and $\tilde U_\nu \times$ a diagonal $U(3)$ matrix.
Without loss of generality we can adopt the following form for $U$:
\be
U=U_e^\dagger U_\nu = \underbrace{ \tilde U_e^\dagger  
{\rm diag}(-e^{- i (\alpha_1 + \alpha_2)}, - e^{-i \alpha_2},1)  \tilde U_\nu }_{= \bar U}
\times {\rm phases}
\label{Ugen}
\ee
where $\tilde U_e$, $\tilde U_\nu$ have the form (\ref{ufi}), the phases $\alpha_1$, $\alpha_2$ run 
from $0$ to $2 \pi$ and 
we have introduced two minus signs in the diagonal matrix for later convenience. 
This expression for $\bar U$ is not due to the Majorana nature of neutrinos and a similar result 
would also hold for quarks.
More technical details on the parametrization (\ref{Ugen}) can be found in the Appendix B.

Assume now that $\tilde U_\nu$ corresponds to bimixing: $s^\nu_{13}=0$,
$s^\nu_{12}=c^\nu_{12}=1/\sqrt{2}$ and $s^{\nu}_{23}=c^{\nu}_{23}=1/\sqrt{2}$. 
Clearly, our discussion holds true irrespectively of the light neutrino spectrum. 
It is anyway instructive to explicitate the mass matrices, 
e.g. in the case of inverted hierarchy 
\be
m_\nu= 
\left(\matrix{0&1&1 \cr 1&0&0\cr1&0&0     } 
\right)\frac{\Delta m^2_{atm}}{\sqrt{2}}~,
\quad 
m_e= V_e  \left(\matrix{ 
m_e e^{-i(\alpha_1 + \alpha_2)}& - s^e_{12} m_e e^{-i \alpha_2}& - s^e_{13} m_e e^{i \delta_e}  \cr 
s^e_{12} m_\mu e^{-i(\alpha_1 + \alpha_2)}& m_\mu e^{-i \alpha_2} & - s^e_{23} m_\mu \cr 
s^e_{13} m_\tau e^{-i(\alpha_1 + \alpha_2 + \delta_e)}& s^e_{23} m_\tau e^{-i \alpha_2} &m_\tau   } 
\right)
\label{}
\ee 
where we have set $\Delta m^2_{sol} = 0$ since, as already mentioned, the 
corrections induced by setting it to the measured value are negligible in the
present discussion.

We then expand $\bar U$ of eq. (\ref{Ugen}) at first order in the small mixings of $\tilde U_e$,
$s^e_{12},$ $s^e_{13}$ and $s^e_{23}$
\footnote{To this approximation any ordering of the three 
small rotations in $U_e$ gives exactly the same results,
and our conclusions are independent on the adopted parametrization.}:
\bea
\bar U_{11}&=&- \frac{e^{-i(\alpha_1 + \alpha_2)}}{\sqrt{2}}
                - \frac{s^e_{12} e^{-i \alpha_2} + s^e_{13} e^{i \delta_e}}{2}\nn\\
\bar U_{12}&=&- \frac{e^{-i(\alpha_1 + \alpha_2)}}{\sqrt{2}}
                + \frac{s^e_{12} e^{-i \alpha_2} + s^e_{13} e^{i \delta_e}}{2}\nn\\
\bar U_{13}&=&  \frac{s^e_{12} e^{-i \alpha_2} - s^e_{13} e^{i \delta_e}}{\sqrt{2}}\nn\\
\bar U_{23}&=& - e^{-i \alpha_2}\frac{ 1 + s^e_{23} e^{i \alpha_2}  }{\sqrt{2}}\nn\\
\bar U_{33}&=&   \frac{1 - s^e_{23} e^{-i \alpha_2} }{\sqrt{2}}~~~.~~~~\label{UeUnBm}
\eea
The smallness of the observed $s_{13}$ implies that both $s^e_{12}$ and $s^e_{13}$ must be 
at most of order $s_{13}$. 
As a consequence, the amount of the deviation of $s_{12}$ from $1/\sqrt{2}$ is 
limited from the fact that it is generically of the same order as $s_{13}$.
Note that, instead, the deviation of the atmospheric angle $s_{23}$ from $1/\sqrt{2}$
is of second order in $s^e_{12}$ and $s^e_{13}$,  
so that it is natural to expect a smaller deviation as observed.
From eqs. (\ref{UeUnBm}) we obtain the following explicit expressions
for the observable quantities
\footnote{Eqs. (\ref{tan23},\ref{deltasol},\ref{ue3}) 
have been independently derived also in Ref. \cite{Frampton}.}:
\bea
\tan^2 \theta_{23} &=& 1 + 4 s^e_{23} \cos(\alpha_2)
\label{tan23}\\
\delta_{sol} \equiv
1-\tan^2 \theta_{12} &=& 2 \sqrt{2} ( s^e_{12} \cos(\alpha_1) + s^e_{13} \cos(\delta_e+\alpha_2+\alpha_1))
\label{deltasol}\\
|U_{e3}| &=&\frac{1}{\sqrt{2}} ( {s^e_{12}}^2 +{s^e_{13}}^2 -2 \cos(\delta_e+\alpha_2) s^e s^e_{13} )^{1/2} 
\label{ue3}\\
\tan \delta &=& \frac{s^e_{12} \sin(\alpha_1) - s^e_{13} \sin(\delta_e+\alpha_2+\alpha_1) }
{s^e_{12} \cos(\alpha_1) - s^e_{13} \cos(\delta_e+\alpha_2+\alpha_1)}~~~~~,
\label{delta}
\eea
to be compared with the experimental data. 
According to \cite{Valle}
the 3-$\sigma$ windows are $|U_{e3}| \le 0.23$ and $0.36 \le \delta_{sol} \le 0.70$.

Notice that the sign of $\delta_{sol}$ is not necessarily positive, 
so that only a part (say half) of the parameter space in principle allowed for the phases 
is selected.
With the correction to $\delta_{sol}$ going in the good direction, 
one roughly expects $|U_{e3}| \sim \delta_{sol}/4 \approx 0.1-0.2$.
Hence, at present it is not excluded that charged lepton mixing can transform a 
bimixing configuration into a realistic one but there are constraints and, 
in order to minimize the impact of those constraints, 
$|U_{e3}|$ must be within a factor of 2 from its present upper limit. 
On the other hand, an upper limit on $|U_{e3}|$ smaller than $\delta_{sol}/4$
would start requiring a fine-tuning. 
Indeed, in order to reduce $|U_{e3}|$ significantly below $0.1-0.2$
a cancellation must be at work in eq. (\ref{ue3}), 
namely $\delta_e+\alpha_2$ should be close to $0$ or $2 \pi$ 
and $s^e_{12}$ and $s^e_{13}$ should be of comparable magnitude.
In addition, to end up with the largest possible $\delta_{sol}/4$, 
eq. (\ref{deltasol}) would also suggest a small value for $\alpha_1$.

\begin{figure}[!h]
\vskip .1 cm
\centerline{~~   
\psfig{file=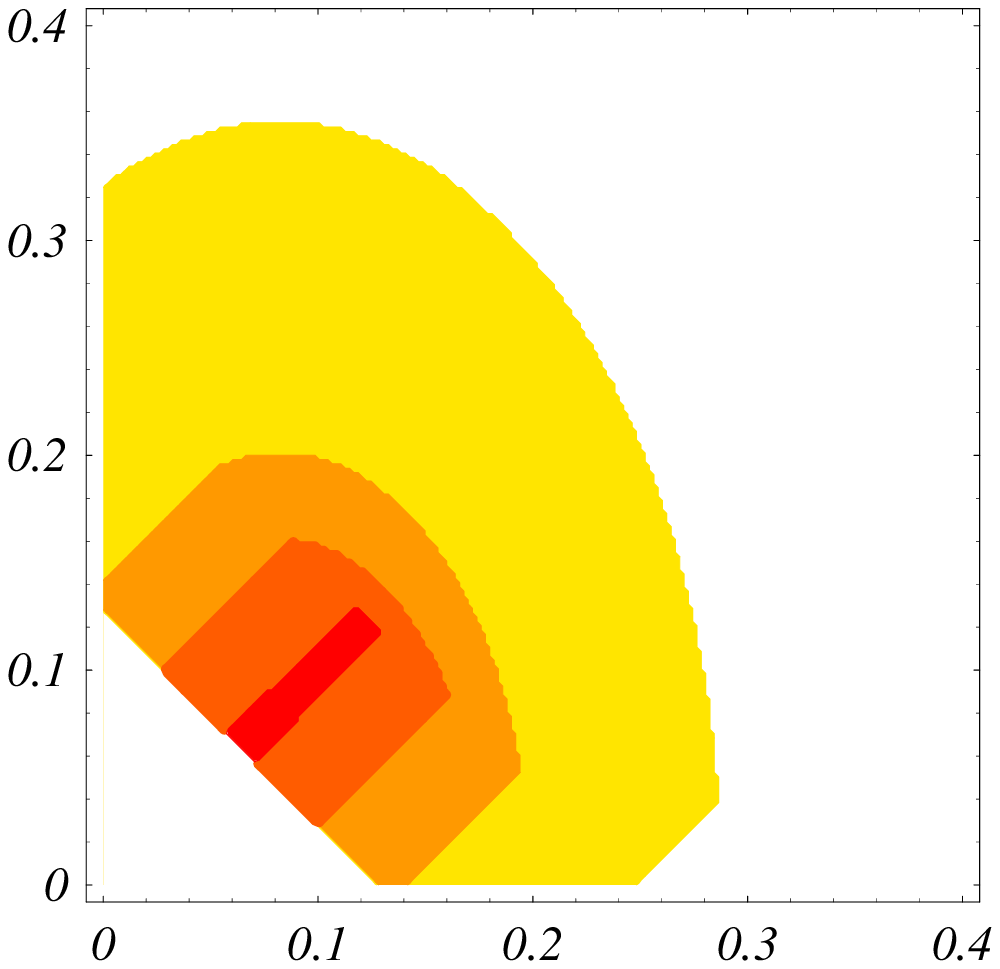,width=0.34 \textwidth}~~
\psfig{file=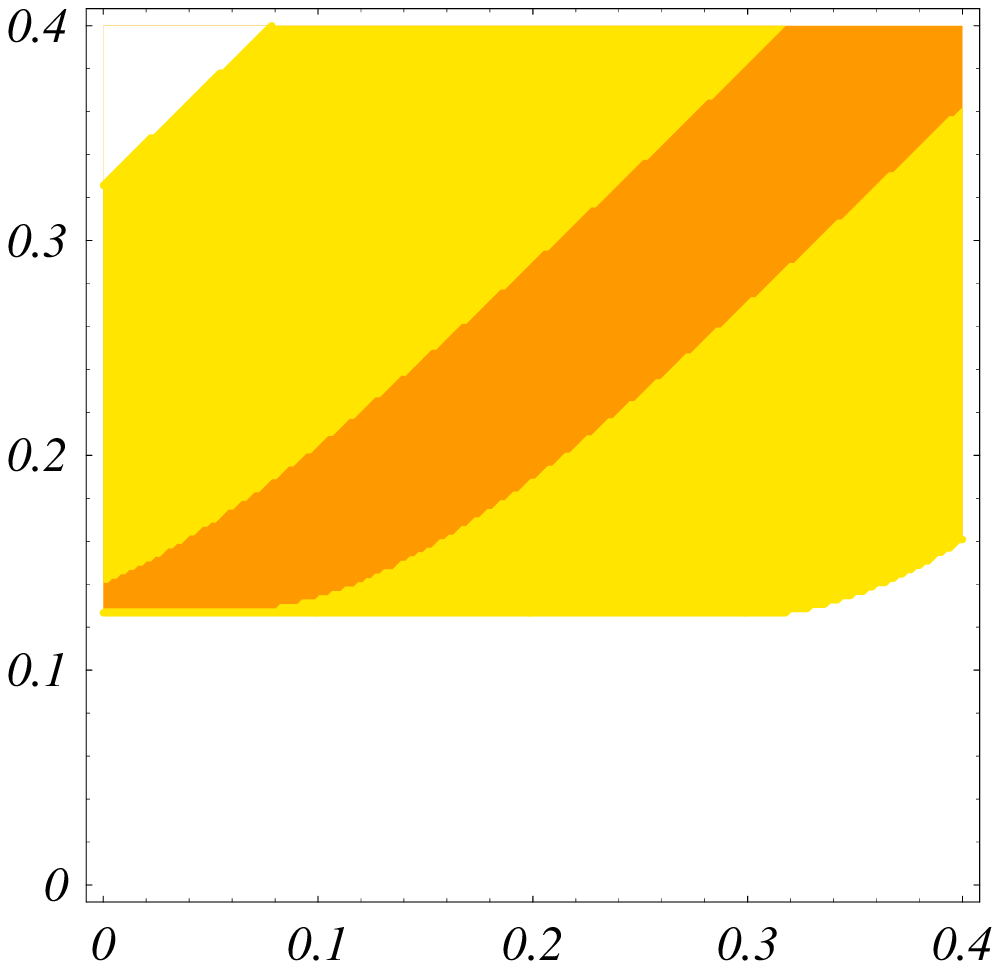,width=0.34 \textwidth}~~
\psfig{file=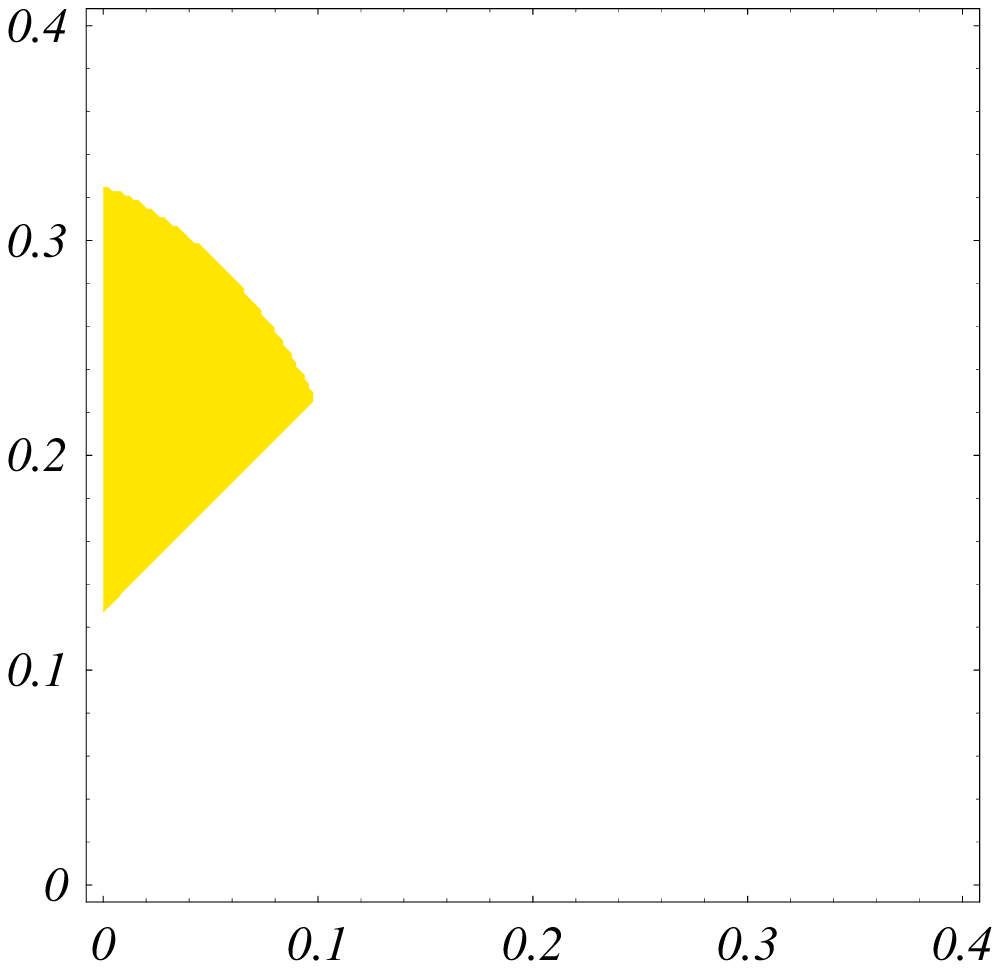,width=0.34 \textwidth} 
\put(-400, 150){ $\alpha_1=0$~~~~~~~~~~~~~~~~~~~~~ 
$\alpha_1=\pi /2$ (or $3 \pi/2$)~~~~~~~~~~~~~~~~~~~~~$\alpha_1=\pi$} 
\put(-360, -10){ $s^e_{12}$ } \put(-200, -10){ $s^e_{12}$ } \put(-50, -10){ $s^e_{12}$ } 
\put(-470, 120){ $s^e_{13}$ } \put(-313, 120){ $s^e_{13}$ }  \put(-156, 120){ $s^e_{13}$ }  
\put(-400, 110){ \footnotesize $.23$ }
\put(-190, 50){\footnotesize $.23$ }
\put(-125, 110){\footnotesize $.23$ }
\put(-420, 75){ \footnotesize $.1$ }
\put(-200, 95){\footnotesize $.1$ }
\put(-425, 60){ \footnotesize $.05$ }
\put(-420, 40){ \footnotesize $.01$ }
}
\caption{Taking an upper bound on $|U_{e3}|$ respectively equal to ${0.23, 0.1, 0.05, 0.01}$,
we show (from yellow to red) the allowed regions of the plane $[s^e_{12}, s^e_{13}]$. We do not display the region outside $s^e_{12}, s^e_{13}\le 0.4$,
where our approximation, linear in $s^e_{12}$ and $s^e_{13}$, breaks down.
Each plot is obtained by setting $\alpha_1$ to a particular value, 
while leaving $\alpha_2 + \delta_e$ free.  
We keep the present 3 $\sigma$ window for $\delta_{sol}$ \cite{Valle}.}
\label{isole}
\end{figure}

The above considerations can be made quantitative by showing, 
for different upper bounds on $|U_{e3}|$,
the points of the plane $[s^e_{12}, s^e_{13}]$ which are compatible with the present
3 $\sigma$ window for the solar angle. 
This is shown in fig. \ref{isole}, where the three plots correspond to different choices
for $\alpha_1$. 
A point in the plane $[s^e_{12}, s^e_{13}]$ is excluded if there is no value 
of $\alpha_2 + \delta_e$ for which (\ref{ue3}) and (\ref{deltasol}) agree with experiment.
Regions in white are those excluded by the present bound on $|U_{e3}|$. 
With increasingly stronger bounds on $|U_{e3}|$, 
the allowed regions, indicated in the plots 
with increasingly darkness, get considerably shrinked.
For $|U_{e3}| \le 0.05$ only $|\alpha_1| < \pi/2$ is allowed.
Notice also that at present the two most natural possibilities $s^e_{12} \gg s^e_{13}$ and
$s^e_{12} \ll s^e_{13}$ are allowed but, with $|U_{e3}| < 0.1$, they are significantly constrained
and with $|U_{e3}| \le 0.05$ ruled out completely. 
Below the latter value for $|U_{e3}|$, a high level of degeneracy between 
$s^e_{12}$ and $s^e_{13}$ together with a small value for 
$\alpha_1$ and $\delta_e + \alpha_2$ are required. 

Summarising, planned improvements in the sensitivity to
$|U_{e3}|$ - which could reach the 0.05 level \cite{future} -,
could have a crucial impact on bimixing models.
They could either disfavour it as unnatural (in the sense that a dynamical
principle or a symmetry acting also on the charged lepton mass matrix would 
have to be invoked) 
or, if $|U_{e3}|$ were to be found, support bimixing models.


\section{Conclusion}

We have considered the possibility that the observed pattern of neutrino mixings is dominantly determined by the charged lepton sector, that is by $U_e$, with a nearly diagonal neutrino matrix
$U_\nu$. Of course, one can always choose an ad hoc basis where this is true: the point is to decide whether this formal choice can be naturally justified in the physical basis where the symmetries of the lagrangian are naturally specified. We find that in presence of two large mixing angles $\theta_{12}$ and $\theta_{23}$ with the third angle $\theta_{13}$ being small, the construction of a natural model with dominance of $U_e$ is made much more difficult than in the case of only the atmospheric angle $\theta_{23}$ large. We have discussed the conditions that must be fulfilled and presented a model where indeed all observed mixings arise from $U_e$. The stated difficulty is reflected in the relatively complicated symmetry structure required and in the fact that we were led to implement in the charged lepton sector a see-saw mechanism as a trick to naturally obtain an approximately vanishing subdeterminant. While the see-saw mechanism is a common and natural occurrence in the neutrino sector, its presence in the charged lepton sector is much more special.

We have also studied the case where the matrix $U_\nu$ is of the bimixing type, 
which is naturally obtained, for example, in inverse hierarchy models with 
a $L_e-L_\mu-L_\tau$ U(1) symmetry, and the observed pattern of mixings is obtained 
by adding corrections from $U_e$. 
We have shown that the smallness of $\theta_{13}$ imposes strong constraints on the 
maximum deviation of the solar angle from maximality, while the deviations of 
the atmospheric angle from the maximal value can be naturally smaller. 
Planned experimental searches for $|U_{e3}|$ could thus have a strong impact in
supporting or disfavouring bimixing.


\section*{Acknowledgements}

We gratefully acknowledge Andrea Romanino for a stimulating 
conversation that attracted our interest on the subject and anticipated 
some of the conclusions that are substanciated here.
I.M. thanks the Dept. of Physics of Rome1 for kind hospitality. 
F.F. and I.M thanks the CERN Theory Division where they were Visitors for periods in summer and 
respectively in fall 2003 and beginning 2004.
This project is partially
supported by the European Programs HPRN-CT-2000-00148 and HPRN-CT-2000-00149.


\section*{Appendix A}

Here we sketch a supersymmetric 
SU(5) grand unified model with a flavour symmetry 
F=U(1)$_{F_0}\times$U(1)$_{F_1}\times$U(1)$_{F_2}\times$U(1)$_{F_3}$.
The field content and the transformation properties under F
are collected in tables 1 and 2.
\\[0.1cm]
{\begin{center}
\begin{tabular}{|c|c|c|c|c|c|c|c|c|c|c|c|c|}   
\hline
&$10_1$ &$10_2$ &$10_3$ 
&$\bar{5}^l_1$ &$\bar{5}^l_2$ &$\bar{5}^l_3$
&$5_H$ &$\bar{5}_H$
&$5_1$ &$5_2$ &$\bar{5}_1$ &$\bar{5}_2$\\
\hline
\hline      
F$_0$ & 4 & 2 & 0 & 0 & 0 & 0 & 0 & 0 & 0 & 0 & 0 & 0 \\
\hline
F$_1$ & 2 & 2 & 2 & 1 & 0 & 0 & 0 & 0 & 2 & 0 & -2 & 0 \\      
\hline
F$_2$ & 2 & 2 & 2 & 0 & 1 & 0 & 0 & 0 & 2 & 0 & -2 & 0 \\      
\hline
F$_3$ & 2 & 2 & 2 & 0 & 0 & 1 & 0 & 0 & 0 & 2 & 0 & -2 \\      
\hline
\end{tabular} 
\end{center}}
\vspace{3mm}
Table 1. Matter fields and flavour charges. 
Fields are denoted by their transformation properties under SU(5). 
Higgs doublets are in $(5_H,\bar{5}_H)$.

{\begin{center}
\begin{tabular}{|c|c|c|c|c|c|c|c|c|}   
\hline
&$\theta_0$  &$\theta_1$  &$\theta_2$  &$\theta_3$ &$\theta_4$
&$\theta_5$ &$\theta_6$ &$\theta_7$\\                       
\hline
\hline      
F$_0$ & -1 & 0 & 0 & 0 & 0 & 0 & 0 & 0 \\
\hline      
F$_1$ & 0 & -2 & 0 & 0 & -3 & 0 & 0 & -4 \\
\hline    
F$_0$ & 0 & 0 & -2 & 0 & 0 & -3 & 0 & -4 \\
\hline      
F$_0$ & 0 & 0 & 0 & -2 & 0 & 0 & -3 & -4 \\
\hline
\end{tabular} 
\end{center}}
\vspace{3mm}
Table 2. Flavon fields and their flavour charges. 

The superpotential is given by:
\be
W=
y_{ij} 10_i 10_j 5_H +
\frac{c_i}{\Lambda}(\bar{5}^l_i 5_H)(\bar{5}^l_i 5_H) +
\eta_{ij} 10_i \bar{5}^l_j \bar{5}_H +
\lambda_{ia} 10_i \bar{5}_a \bar{5}_H +
\mu_{aj} 5_a \bar{5}^l_j +
M_a 5_a \bar{5}_a~~~~~,
\ee      
where $\Lambda$ denotes a large mass scale, close to the cut-off
of the theory, and all the coupling constants are proportional to powers
of $\langle\theta_i\rangle/\Lambda$ that can be read off from 
tables 1 and 2. Here for simplicity we assume that all flavon fields 
acquire a similar VEV, $\langle\theta_i\rangle/\Lambda\approx\lambda$.
In this case we have:
\be
c_i=\hat{c}_i \lambda~~,~~~~~
\eta_{ij}=\hat\eta_{ij} \lambda^{(9-2i)}~~,~~~~~
\lambda_{ia}=\hat\lambda_{ia} \lambda^{(6-2i+a)}~~,~~~~~
\mu_{aj}=\hat\mu_{aj} \lambda^{(3-a)}~~~~~,
\ee
where hatted quantities (and $M_a$) do not depend on $\lambda$.
We also have $\mu_{13}=\mu_{21}=\mu_{22}=0$.

We get for the up-type quarks a mass matrix of the form:
\be 
m_u=
\left[\matrix{
\lambda^8&\lambda^6&\lambda^4\cr 
\lambda^6&\lambda^4&\lambda^2\cr 
\lambda^4&\lambda^2&1} 
\right] \lambda \langle 5_H \rangle~~~~~. 
\label{mumod}
\ee
After eliminating the heavy degrees of freedom, in the regime
$1>|\mu_{aj}/M_a|>|\eta_{ia}/\lambda_{ij}|$, the mass matrix for charged leptons 
is dominated by the see-saw contribution and is given by:
\be
m_e= -
\left[\matrix{
\dd\frac{\hat\lambda_{11}\hat\mu_{11}}{M_1} \lambda^4&
\dd\frac{\hat\lambda_{11}\hat\mu_{12}}{M_1} \lambda^4&
\dd\frac{\hat\lambda_{12}\hat\mu_{23}}{M_2} \lambda^4\cr
\dd\frac{\hat\lambda_{21}\hat\mu_{11}}{M_1} \lambda^2&
\dd\frac{\hat\lambda_{21}\hat\mu_{12}}{M_1} \lambda^2&
\dd\frac{\hat\lambda_{22}\hat\mu_{23}}{M_2} \lambda^2\cr
\dd\frac{\hat\lambda_{31}\hat\mu_{11}}{M_1} &
\dd\frac{\hat\lambda_{31}\hat\mu_{12}}{M_1} &
\dd\frac{\hat\lambda_{32}\hat\mu_{23}}{M_2}} 
\right ]\lambda^3\langle \bar{5}_H \rangle~~~~~.
\label{memod}
\ee
We see that the first and second column are dominated by the
$M_1$ contribution, thus producing the desired vanishing determinant.
Actually, in the considered regime, the first two column are
exactly proportional and the electron mass vanishes. This can be
corrected by sub-dominant terms, arising for instance from the 
neglected contributions.
The neutrino mass matrix is diagonal, with masses given by
\be
m_i=\hat{c}_i
\frac{\langle 5_H \rangle^2\lambda}{\Lambda}~~~~~.
\ee

If we assume dimensionless coefficients 
$\hat\eta$ and $\hat\lambda$ of order one, the condition $1>|\mu_{aj}/M_a|>|\eta_{ia}/\lambda_{ij}|$ 
requires to choose the dimensionful parameters $\hat\mu_{ai}$ and $M_a$
in the window  $1<\hat\mu_{ai}/M_a<1/\lambda^2$.


\section*{Appendix B}

We show here that (\ref{Ugen}) is a general way of writing $U$.
Similar parametrizations have also been used in \cite{King, par}.
Starting from the basis of the (unknown) flavour symmetry
\be
{\cal L} = \nu^T m_\nu \nu + e^{cT} m_e e + e^\dagger \nu W
\ee 
then 
\be
m_\nu = U_\nu^* m_\nu^{diag} U_\nu^{\dagger} ~~~~~~ m_e = V_e m_e^{diag} U_e^{\dagger} 
\ee
where $diag$ stands for a diagonal matrix with real non-negative elements and $U_\nu$, $U_e$, $V_e$
are unitary matrices. The MNS mixing matrix $U$ is then given by $U=U_e^\dagger U_\nu$.
We are going to exploit the fact that any unitary matrix can in general be written unambiguously as:
\be
U= e^{i \phi_0} \underbrace{ {\rm diag}(e^{i(\phi_1 + \phi_2)}, e^{i \phi_2}, 1)}_{\equiv \Phi} \tilde U 
                \underbrace{ {\rm diag}(e^{i(\phi_3 + \phi_4)}, e^{i \phi_4}, 1)}_{\equiv \Phi'} ~,
\ee
where $\phi_i$ (i=0,...,4) run from $0$ to $2 \pi$ and $\tilde U$ is the standard parameterization for the CKM
mixing matrix, namely
\be
\tilde U~=~ 
\left(\matrix{1&0&0 \cr 0&c_{23}&s_{23}\cr0&-s_{23}&c_{23}     } 
\right)
\left(\matrix{c_{13}&0&s_{13}e^{i\delta} \cr 0&1&0\cr -s_{13}e^{-i\delta}&0&c_{13}     } 
\right)
\left(\matrix{c_{12}&s_{12}&0 \cr -s_{12}&c_{12}&0\cr 0&0&1     } 
\right)~~~,
\label{uffi}
\ee 
where all the mixing angles belong to the first quadrant and $\delta$ to $[0,2 \pi]$.
If we adopt such a parameterization for both $U_e$ and $U_\nu$ we obtain, 
in the basis where mass matrices are diagonal:
\be
U = e^{i (\phi^\nu_0 -\phi^e_0)}~ \Phi'^*_e ~\tilde U_e^\dagger ~\Phi^*_e \Phi_\nu ~ \tilde U_\nu^\dagger ~ \Phi'_\nu  ~~.
\label{step1}
\ee
Now, by redefining properly left and right-handed charged lepton fields, 
we can get rid of all the phases at the left of $\tilde U_e^\dagger$.
The phases in between $\tilde U_e^\dagger$ and $\tilde U_\nu$ cannot be eliminated, 
but we can always define their product as:
\be
\Phi^*_e \Phi_\nu \equiv {\rm diag}(-e^{-i(\alpha_1 + \alpha_2)}, -e^{-i \alpha_2}, 1)~~.  
\ee
Also the phases on the right of $\tilde U_\nu$ cannot be eliminated and contribute to the Majorana phases.
As a result of all the above redefinitions, eq. (\ref{step1}) becomes:
\be
U = \tilde U_e^\dagger ~{\rm diag}(-e^{-i(\alpha_1 + \alpha_2)}, -e^{-i \alpha_2}, 1) ~ 
\tilde U_\nu^\dagger ~\Phi'_\nu~~,
\label{step2}
\ee 
which is precisely the parameterization used in the text (see eq. (\ref{Ugen})).

\vfill
\newpage


\end{document}